\documentstyle[12pt]{article}
\def\beq{\begin{equation}}
\def\eeq{\end{equation}}
\def\bea{\begin{eqnarray}}
\def\eea{\end{eqnarray}}

\catcode`\@=11
\def\marginnote#1{}
\def\ifmath#1{\relax\ifmmode #1\else $#1$\fi}

\def\bold#1{\setbox0=\hbox{$#1$}%
     \kern-.025em\copy0\kern-\wd0
     \kern.05em\copy0\kern-\wd0
     \kern-.025em\raise.0433em\box0 }

\def\GENITEM#1;#2{\par\vskip6pt \hangafter=0 \hangindent=#1
   \Textindent{$ #2$ }\ignorespaces}

\newcount\hour
\newcount\minute
\newtoks\amorpm
\hour=\time\divide\hour by60
\minute=\time{\multiply\hour by60 \global\advance\minute by-
\hour}
\edef\standardtime{{\ifnum\hour<12 \global\amorpm={am}%
    \else\global\amorpm={pm}\advance\hour by-12 \fi
    \ifnum\hour=0 \hour=12 \fi
    \number\hour:\ifnum\minute<100\fi\number\minute\the\amorpm}}
\edef\militarytime{\number\hour:\ifnum\minute<100\fi\number\minute}
\def\draftlabel#1{{\@bsphack\if@filesw {\let\thepage\relax
  \xdef\@gtempa{\write\@auxout{\string
    \newlabel{#1}{{\@currentlabel}{\thepage}}}}}\@gtempa
    \if@nobreak \ifvmode\nobreak\fi\fi\fi\@esphack}
     \gdef\@eqnlabel{#1}}
\def\@eqnlabel{}
\def\@vacuum{}
\def\draftmarginnote#1{\marginpar{\raggedright\scriptsize\tt#1}}
\def\draft{\oddsidemargin -.5truein
        \def\@oddfoot{\sl preliminary draft \hfil
        \rm\thepage\hfil\sl\today\quad\militarytime}
        \let\@evenfoot\@oddfoot \overfullrule 3pt
        \let\label=\draftlabel
        \let\marginnote=\draftmarginnote

\def\@eqnnum{(\theequation)\rlap{\kern\marginparsep\tt\@eqnlabel}%
\global\let\@eqnlabel\@vacuum}  }
\def\preprint{\twocolumn\sloppy\flushbottom\parindent 1em
        \leftmargini 2em\leftmarginv .5em\leftmarginvi .5em
        \oddsidemargin -.5in    \evensidemargin -.5in
        \columnsep 15mm \footheight 0pt
        \textwidth 250mmin      \topmargin  -.4in
        \headheight 12pt \topskip .4in
        \textheight 175mm
        \footskip 0pt

\def\@oddhead{\thepage\hfil\addtocounter{page}{1}\thepage}
        \let\@evenhead\@oddhead \def\@oddfoot{} \def\@evenfoot{}
}
\def\titlepage{\@restonecolfalse\if@twocolumn\@restonecoltrue\o
necolumn
     \else \newpage \fi \thispagestyle{empty}\c@page\z@
        \def\thefootnote{\fnsymbol{footnote}} }
\def\endtitlepage{\if@restonecol\twocolumn \else  \fi
        \def\thefootnote{\arabic{footnote}}
        \setcounter{footnote}{0}}  
\catcode`@=12
\relax
\def\be{\begin{equation}}
\def\ee{\end{equation}}
\def\bea{\begin{eqnarray}}
\def\eea{\end{eqnarray}}
\def\simlt{\stackrel{<}{{}_\sim}}
\def\simgt{\stackrel{>}{{}_\sim}}

\def\mst11{m_{\;\widetilde{t}_{1}}}

\def\mst22{m_{\;\widetilde{t}_{2}}}
\def\mst12{m_{\;\widetilde{t}_{1,2}}}

\def\msb11{m_{\;\widetilde{b}_{1}}}
\def\msb22{m_{\;\widetilde{b}_{2}}}
\def\msb12{m_{\;\widetilde{b}_{1,2}}}

\def\mwidetilde2{\widetilde{m}^{2}}

\relax

\begin{document}
\input epsf

\topmargin-2.5cm
%
\begin{titlepage}
\begin{flushright}
CERN-TH/2001-082\\
hep--ph/0103248 \\
\end{flushright}
\vskip 0.1in
\begin{center}
{\Large\bf  The Cosmological Moduli Problem and  Preheating}

\vskip .5in
{\large\bf G.F. Giudice$^{1}$},
{\large \bf A. Riotto$^{2}$},
{\large and}
{\large\bf I.I. Tkachev$^{1}$}

\vskip0.7cm
$^{1}$CERN Theoretical Physics Division,

\vskip 0.2cm

CH-1211 Geneva 23, Switzerland

\vskip 0.5cm

$^{2}$INFN, Sezione di Padova,                      

\vskip 0.2cm

Via Marzolo 8, Padova I-35131, Italy

\end{center}
\vskip 1cm
\begin{center}
{\bf Abstract}
\end{center}
\begin{quote}

Many models of supersymmetry breaking, in the context of either supergravity 
or superstring theories, predict the presence of particles with 
Planck-suppressed couplings and masses around the weak scale. These particles 
are generically called moduli. The excessive production of moduli in the early 
Universe jeopardizes the successful predictions of nucleosynthesis. In this 
paper we show that the efficient generation of these dangerous relics is an
unescapable consequence of a wide variety of inflationary models which have a 
preheating stage. Moduli are generated as coherent states in a novel way which
differs from the usual production mechanism during parametric resonance. The 
corresponding limits on the reheating temperature are  often very tight and 
more severe than the bound of $10^9$ GeV coming from the production of moduli 
via thermal scatterings during reheating.

\end{quote}
\vskip1.cm
\begin{flushleft}
March 2001 \\
\end{flushleft}

\end{titlepage}
\setcounter{footnote}{0}
\setcounter{page}{0}
\newpage
%
\baselineskip=18pt
\noindent

\section{Introduction}

In  supergravity and (super)string models 
there  exists a plethora of 
scalar and fermionic fields (we will loosely call them moduli 
$\Phi$) with masses of the 
order of the weak scale and 
gravitational-strength couplings to 
ordinary matter. 
If produced in the early Universe, such quanta   behave 
like nonrelativistic matter and decay at very late times, eventually 
dominating the energy of the
 Universe until it is too late for nucleosynthesis to 
occur. This poses a serious cosmological problem~\cite{moduli}.
Typical examples of these dangerous relics 
are the spin-3/2 gravitino (the supersymmetric partner of the graviton), the scalar moduli  
which
parametrize  supersymmetric flat directions in moduli space and seem almost 
ubiquitous in string theory, and their fermionic superpartners.

The slow decay rate of the $\Phi$-particles
is the essential source of cosmological problems because 
the  decay products of these relics  will destroy the $^4$He and 
D nuclei by photodissociation, 
and thus successful nucleosynthesis predictions \cite{nucleo,ellis}. The most 
stringent bound 
comes from the resulting 
overproduction of D $+$ $^3$He; this requires that the 
$\Phi$-abundance 
 relative to the entropy density $s$ at the time of reheating after 
inflation \cite{lr} should satisfy~\cite{kaw}
\begin{equation}
\label{lll}
\frac{n_{\Phi}}{s}\simlt 10^{-12},
\end{equation}
where the 
exact bound depends upon the mass $m_\Phi$. 

The dangerous  moduli  can be produced in the early Universe in a number
 of ways\footnote{Here we will study only production of moduli
quanta and we do not consider the possibility of moduli oscillations
around their minimum.}. They may be generated by 
thermal scatterings in the plasma during the process of reheating after 
inflation when the energy density of the Universe gets converted into
radiation. The  number  density $n_\Phi$             
at thermalization is estimated to be of the order of $n_\Phi/s\simeq 10^{-2}\:
(T_r/M_{{\rm Pl}})$, 
where $T_r$ is the final reheating temperature and $M_{{\rm Pl}}=1.2 \times
10^{19}$ GeV is the Planck mass.
This gives a stringent  upper 
bound on the reheating temperature after inflation  of about $10^{9}$
GeV \cite{ellis}.

The generation of the dangerous relics may also proceed through
non-thermal effects. Scalar moduli fields may be copiously 
created by the classical gravitational effects on the vacuum state
between the end of inflation and the beginning of the 
matter/radiation phase \cite{glv,gkr1,fel}. In such a case, the 
 upper bound on the reheating temperature is very dependent on the inflationary
model, but it
may be  as low as 100 GeV. 
Another possibility is represented 
by the parametric excitation of dangerous relics occuring 
after inflation because of  the rapid oscillations of the inflaton field(s).
Gravitino production is an interesting example of this kind 
\cite{gkr1,l1,gkr2,l2} since the 
non-thermal generation of gravitinos in the early Universe can be 
extremely efficient and overcome the
thermal production by several orders of magnitude\footnote{
Recently, this statement has been criticized in ref.~\cite{nps}, where it
was shown that, in a simple Polonyi model, gravitino production is
suppressed. Undoubtedly, as emphasized in refs.~\cite{gkr1,l1,gkr2,l2},
the 
non-thermal
gravitino production depends sensitively on the assumptions on the 
inflationary model and its couplings to the supersymmetry-breaking sector.
Non-thermal production of gravitinos is expected to be 
significant and to pose a cosmological problem if -- after inflation --
there is a strong mixing between
the inflaton sector and the supersymmetry-breaking sector (of the present
vacuum). As it will become clear  in the present paper, this is often the
case because, during the first few inflaton
oscillations when the value of the inflaton field is close to $M_{\rm Pl}$,  
generic Planck mass-suppressed interactions can give a large contribution 
to such mixing.}.

The scope of this  paper is to  show that the abundant 
production of dangerous relics
is an unavoidable consequence of a period of preheating \cite{pre}
 after inflation. The corresponding limits on the reheating temperature
are often very tight and more severe than the bound of $10^9$ GeV
coming from the production of moduli via thermal scatterings during
reheating.

During the first stage of preheating 
nonperturbative quantum effects lead to an extremely
effective dissipational dynamics and explosive particle production.
Particles coupled to the inflaton field 
can be created in a broad parametric resonance with a fraction of
the energy stored in the form of coherent inflaton oscillations at the end
of inflation released after a few oscillations. 
This rapid transfer of energy density populates the Universe with
quanta in  highly nonthermal states which
can be viewed  either as classical waves travelling through the Universe 
\cite{th1} or as quantum particles in states with large
occupation numbers. In the second stage of preheating, called semiclassical
thermalization \cite{th1}, rescatterings of the produced fluctuations smear
out the resonance peaks in the power spectra and lead to a slowly evolving
state in which the power spectra are smooth \cite{th1,th2,th3}. The system
begins to exibit a chaotic behaviour characteristic of a classical non-linear
system with many degrees of freedom. In the course of the subsequent slow
evolution, the power spectra propagate to larger momenta, 
eventually leading to a fully thermalized state.

In supersymmetric and superstring theories coupling 
constants and masses  appearing in the Lagrangian have to be thought as
functions of the dimensionless ratio $\Phi/M_{{\rm Pl}}$. For instance, a
generic
coupling constant $h$  is in fact a function of the
scalar moduli
\begin{equation}
h(\Phi)=h\left(\frac{\langle\Phi\rangle}{M_{{\rm Pl}}} 
\right)\left(1 +c\frac{\delta\Phi}{M_{{\rm Pl}}}
+\cdots\right),
\end{equation}
where $c$ is a coefficient usually of order unity and
$\delta\Phi=\Phi-\langle\Phi\rangle$.
 This expansion introduces a gravitational
coupling between the scalar moduli  and the matter fields. If the
underlying theory
is supersymmetric, similar couplings of the
 scalar  field $\Phi$ and its fermionic superpartner to
 the matter fields will arise in the
superpotential or in the K\"ahler potential.

Since moduli fields couple
to any form of matter at least gravitationally,
the production is present  in a wide variety of
inflationary models which have a preheating stage. In general, the 
equation of 
motion for the (quantum) field $\delta\Phi$ during preheating is
\begin{equation}
\Box\delta\Phi + m^2 \, \delta\Phi = \sum_i h_i J_i \; ,
\end{equation}
where the currents $J_i$ are  functions of the (classical) fields
created at preheating and  have interaction vertices
parametrized by the coupling constants $h_i$. In a cosmological setting
particle creation in a time-varying classical background 
is usually  caused by the time dependence
of the effective mass of the field (consider, for instance, parametric
resonance
or gravitational creation of particles). As a result, particles
are created in a squeezed state. On the contrary, creation of moduli is
caused
by the non-zero classical current and therefore they 
are created as coherent states. 
{}From a quantum mechanical point of view,  this
production mechanism
is novel and differs from the more traditionally studied
generation  mechanism of squeezed states. 
Furthermore, the moduli production does  not depend
upon the particular properties of the modulus unlike the gravitino
production
\cite{gkr1,l1,gkr2,l2} and it  is true both for scalar and fermionic
moduli. 

In this paper we perform a calculation of moduli production during
the preheating stage in various inflationary models. In sect.~2 we consider
the cases of chaotic inflation with 
quartic and quadratic inflaton potentials, with and without
couplings of the inflaton to a new scalar field $X$. The case of hybrid
inflation is discussed in sect.~3, while sect.~4 contains a study of the
production at preheating of modulinos, light fermions with only
gravitational-strength couplings. In sect.~5 we summarize our
results.   

\section{Moduli Production at Preheating in Chaotic Inflation}

Let us first consider a 
simple model of chaotic inflation with quartic 
potential
$V(\phi)=\frac{\lambda}{4}\phi^4$. As discussed in sect.~1,
we expect 
a generic coupling of the scalar 
moduli quanta $\delta\Phi$ with the inflaton field
$\phi$ of the form $\lambda J\delta\Phi$, where $J \equiv c 
\phi^4/(4M_{{\rm Pl}})$. Let us denote
the amplitude of the oscillating inflaton field as $\phi_0$, while the 
frequency of the oscillations is  $\sim \sqrt{\lambda}\phi_0$, where
$\phi_0$ is the zero-momentum mode of $\phi$.

In this model fluctuations of the inflaton
field do not grow during chaotization (they actually decrease
due to the redshift) so the most efficient production of moduli
fields take place at the end of the resonance phase.
As we noted, one can describe the states we are considering either as a 
collection of interacting classical waves or as a collection of particles
in modes with large occupation numbers. 
Therefore at late times it is possible to treat the system classically
and to solve the relevant equations of motion on a lattice. The initial
conditions
for the classical problem are found solving the preceding quantum
evolution
by means of the appropriate Bogolyubov transformations (for more details
see Ref.~\cite{th1}).

The results of the numerical integration on a $128^3$ lattice  
for $\lambda=10^{-12}$ are summarized in Fig. 1,
where we plot the
values of $\langle\phi^2\rangle$, $\langle\Phi^2\rangle$ and
of the inflaton zero  mode squared $\phi^2_0$ as a function of the 
conformal time \cite{th1} and in units $M^2_{{\rm Pl}}$.
The field variance is given by $\langle\phi^2\rangle =[(2\pi)^3{\cal V}]^{-1}
\int d^3{\bf k} \phi_{\bf k}^2$, where ${\cal V}$ 
is the volume and $\phi_{\bf k}$ is the Fourier
${\bf k}$-mode of the field $\phi$. Here and in the following we have taken 
$c=1$.  Time and
particle momenta 
are given in units of $\sqrt{\lambda} \phi_0(0)$, where $\phi_0(0)$ is the 
initial value of the zero inflaton mode at the end of inflation. The time
$t=0$, when inflation ends,  is chosen as the moment of the first extremum
of $\phi(t)$.
The maximum value of 
$\langle\phi^2\rangle$ is achieved at the time $t_r$, which marks the
end of the resonance phase, and, as shown in Fig.~1 
is about $10^{-7}\;M^2_{{\rm Pl}}$, 
corresponding to $\langle\Phi^2\rangle\sim 10^{-15}\;M^2_{{\rm Pl}}$.
At the same time the zero mode is about $\phi_0(t_r)\sim
10^{-3}\;M_{{\rm Pl}}$. 

\begin{figure}
\begin{center}
\leavevmode\epsfxsize=5.5in\epsfbox{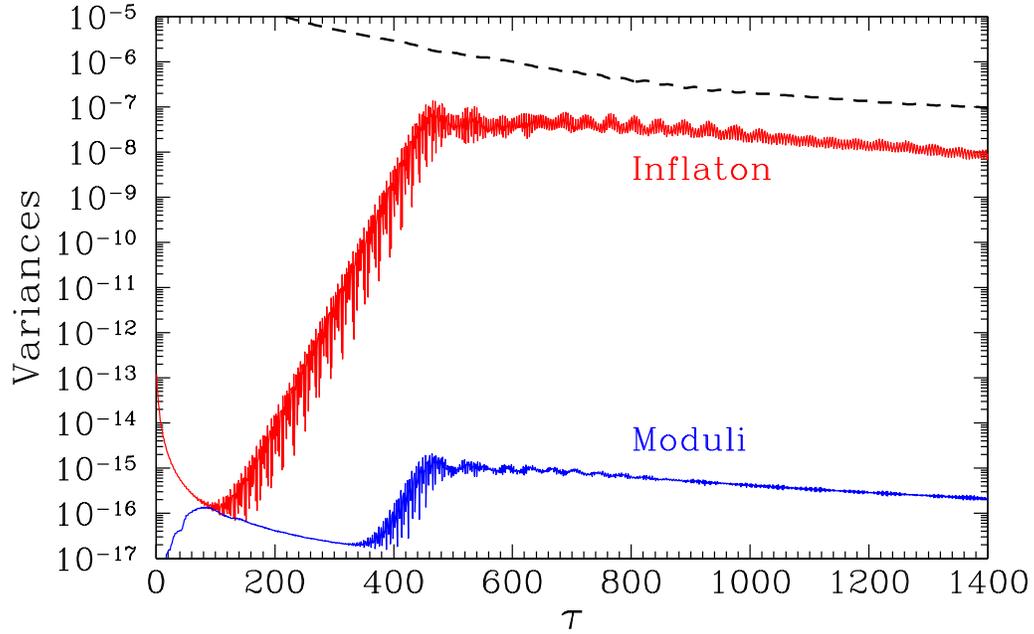}
\end{center}
\caption{The variances of the inflaton and the moduli fields in units of
$M_{\rm Pl}^2$ in a model with inflaton potential $V=\frac{\lambda}{4}
\phi^4$ and moduli interaction $c\lambda \phi^4\delta \Phi /(4M_{\rm Pl})$,
with $\lambda=10^{-12}$ and $c=1$.
The dashed line shows the
amplitude square of the inflaton field 
oscillations in the same units.}
\label{fig:lp4}
\end{figure}

We will now present some analytical estimates which approximately reproduce
the numerical results of Fig.~1. 
To obtain
the intensity of the moduli gravitational production which
accompanies the creation and annihilation of the fluctuations generated
during preheating, we first express the solution of the modulus equation
of motion in terms of the retarded Green function
\begin{equation}
\Phi (x)=\Phi_{in} (x)+\lambda \int d^4x' G_{ret}(x-x')J(x').
\end{equation}
Here $\Phi_{in}$ is the initial configuration at time $t=0$ and
the retarded Green function is given by
\begin{equation}
G_{ret}(x-x')=\theta(t-t') \int \frac{d^3 {\bf k}}{(2\pi)^3}
\frac{\sin \omega_k(t-t')}{\omega_k}e^{i{\vec{\bf k}}\cdot{\vec{\bf x}}}.
\end{equation}
In the case under consideration, 
the source term is given by $J=c\phi^4/(4M_{\rm Pl})$. Expanding 
the fields in Fourier modes, performing the integral
over spatial coordinates, and taking $\phi_0\gg \delta \phi_{\bf k}$
(justified by the numerical results of Fig.~1), we obtain
\begin{equation}
\label{green}
\Phi_{{\bf k}}(t)=\frac{c\lambda}{M_{{\rm Pl}}}\int_0^t\;
dt^\prime\;\frac{\sin\left[\omega_k(t-t^\prime)\right]}{\omega_k}
\phi_0^3(t^\prime)\;\delta
\phi_{{\bf k}}(t^\prime),
\end{equation}

A crucial point is that the effect is linear in the
inflaton
fluctuations $\delta\phi_{{\bf k}}$. Therefore, the phenomenon of moduli
production is due to the  oscillations (similar to oscillations of neutrino
species in a medium) of the inflaton quanta into moduli in presence of the
inflaton background $\phi_0$,  rather than
to real rescatterings. From a pure
quantum mechanical point of view, scalar moduli are generated
as coherent states and not as squeezed states as in the usual
parametric resonance. 

 The fluctuations
$\delta\phi_{{\bf k}}$ grow exponentially during the resonance phase
as $e^{\mu t}$ where  $2\mu\simeq 0.07\sqrt{\lambda}\phi_0(0)$~\cite{th1}
and  $\phi_0(0)\simeq 0.3 \;M_{{\rm Pl}}$ is the initial value
of the inflaton field at the end of inflation.
Therefore the integral in eq.~(\ref{green}) can be approximately
evaluated by assuming that the exponential determines the full time
dependence of the integrand. Then 
we    obtain
that the maximum value of the variance of the modulus field, achieved
at the end of the resonance phase,  is given by
\begin{equation}
\label{es}
\langle\Phi^2\rangle\simeq \frac{c^2\lambda^2~
\phi^6_0(t_r)~\langle\phi^2\rangle}
{\omega_k^2(t_r)\mu^2
M^2_{\rm Pl}}\simeq 10^4 c^2\frac{\phi^4_0(t_r)}{M^4_{\rm Pl}}
~\langle\phi^2\rangle .
\end{equation}
Here we have used $\omega_k(t_r)\sim \sqrt{\lambda}\phi_0(t_r)$
as the typical frequency at the 
end of the resonance phase.
Taking from Fig.~1 the values $\phi_0\sim 10^{-3}M_{\rm Pl}$
and $\langle\phi^2\rangle \sim 10^{-7}M^2_{\rm Pl}$ at $t=t_r$, eq.~(\ref{es})
gives $\langle\Phi^2\rangle \sim c^2 10^{-15}M^2_{\rm Pl}$, in very good
agreement with the numerical result. Notice that the dependence on $\lambda$
dropped out from eq.~(\ref{es}). Our numerical calculation confirms that
the maximum of $\langle\Phi^2\rangle$ is independent of the coupling
$\lambda$ and it scales quadratically with $c$. 

Since it is a good approximation to assume that the dominant particle
production occurs around the time $t_r$, we find
\begin{equation}
n_\Phi\sim \omega_k(t_r)\langle \Phi^2\rangle~~ {\rm and}~~
n_\phi\sim \omega_k(t_r)\langle \phi^2\rangle .
\end{equation}
Using eq.~(\ref{es}) we obtain 
\begin{equation}
\frac{n_\Phi}{n_\phi}\sim 10^{-8} c^2\left[\frac{\phi_0(t_r)}{10^{-3}
M_{\rm Pl}}
\right]^4
\end{equation}
 at the end of the resonance phase.

The  ratio of the number density of moduli 
in units of the entropy density $s\sim \rho^{3/4}\sim
\lambda^{3/4}\phi_0^3(t_r)$ 
is given by
\begin{equation}
\frac{n_{\Phi}}{s}
\sim 10^{-6}c^2\left(\frac{10^{-13}}{\lambda}\right)^{1/4}
\left( \frac{\phi_0^2(t_r)}{10^{-6}M^2_{\rm Pl}}\right)
\left(\frac{\langle \phi^2\rangle}
{10^{-7}M^2_{\rm Pl}} \right) .
\label{nphis}
\end{equation}
For $\lambda =10^{-13}$ (as fixed by the COBE normalization),
this results contradicts 
the bound (\ref{lll}) by about six orders of magnitude (for $c$ of
order unity)
when the energy density in the scalar field 
is transferred to the energy density of a hot gas of relativistic particles.
Notice  that the ratio $\frac{n_{\Phi}}{s}$ does not depend on
the time of thermalization, because both $n_{\Phi}$ and $V^{3/4}$ 
vary with the scale factor $a$ 
as $a^{-3}$.

In this calculation (and throughout all this paper) we have made 
the implicit hypothesis that the
moduli fields are light during the preheating stage. However, the same generic expansion in eq. (2) is expected to
provide the moduli fields with a mass of the order of  the Hubble rate $H$ during the preheating stage.
Consider the simple model of chaotic inflation with potential $V(\phi)=\frac{\lambda}{4}\phi^4$. The
expansion of the coupling $\lambda(\Phi/M_{{\rm Pl}})$ at the quadratic order in 
$\Phi/M_{{\rm Pl}}$ leads to a term
in the
Lagrangian of the form $\sim \lambda(\delta\Phi/M_{{\rm Pl}})^2\phi^4\propto H^2(\delta\Phi)^2$. Nevertheless,  
our  numerical and analytical estimates are not affected by treating the moduli as light states. This is because
moduli fields are generically produced with frequencies $\omega$ much larger than the Hubble rate 
($\omega\sim \sqrt{\lambda}\phi_0(t_r)\gg H(t_r)$ in  the chaotic model $\lambda\phi^4$)
and
kinematically they may be regarded as
massless. 
This does not come as a surprise since it is known that 
superheavy particles may be easily
generated at preheating \cite{s1}.

\begin{figure}
\begin{center}
\leavevmode\epsfxsize=5.5in\epsfbox{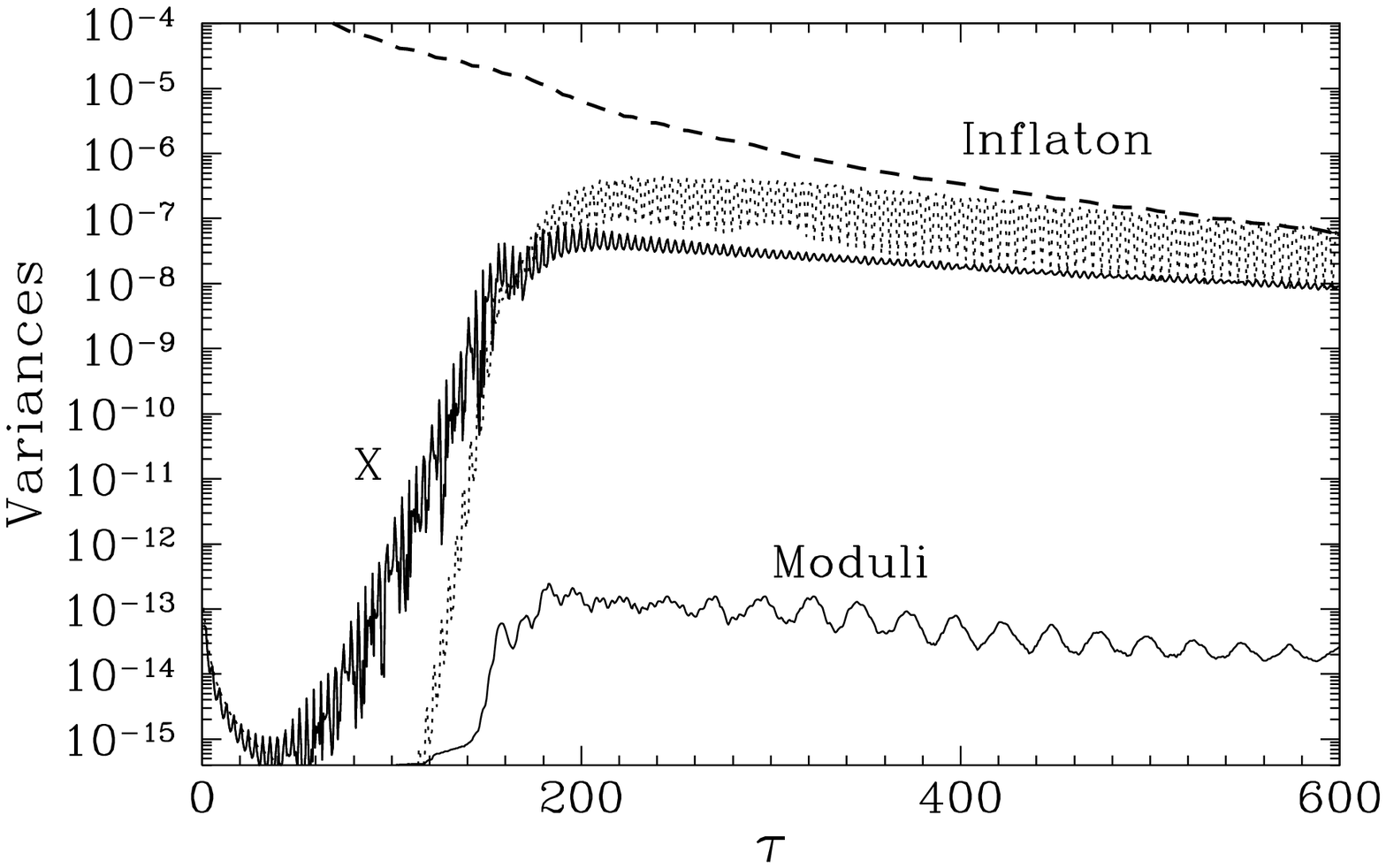}
\end{center}
\caption{The variances of the inflaton, $X$ and moduli fields in units of
$M_{\rm Pl}^2$ in a model with potential $V=\frac{\lambda}{4}
\phi^4+\frac{g^2}{2}\phi^2X^2$ and moduli interaction $c\lambda 
\phi^4\delta \Phi /(4M_{\rm Pl})$,
with $\lambda=10^{-12}$, $q\equiv g^2/(4\lambda)=25$ and $c=1$.
The dashed line shows the
amplitude square of the inflaton field 
oscillations in the same units.}
\label{fig:c2l}
\end{figure}

A simple variation of the model considered so far is the one in which a
massless inflaton field is coupled to another field $X$, $V=\frac{1}{4}
\lambda \phi^4+ \frac{1}{2}g^2\phi^2 X^2$.
The oscillating
field $\phi$  amplifies fluctuations of the field $X$ via
parametric resonance. The strength of the resonance can be parametrized by
the
$q$-parameter 
$q \equiv g^2/4\lambda$. Fluctuations of the inflaton field grow rapidly
because of the back-reaction and the energy of the inflaton zero mode
is quickly released into quanta of the $X$-field and the inflaton field.

For $q=25$, we numerically find that the maximum value of $\langle \phi^2
\rangle$ is
about $5\times 10^{-7} M^2_{{\rm Pl}}$ while $\langle X^2 \rangle$ 
is about one order of magnitude smaller, see Fig. \ref{fig:c2l}. 
The resonance stage
ends when  $\phi^2_0(t_r)\sim 5\times 10^{-6}\;M^2_{{\rm Pl}}\gg \langle
\phi^2\rangle$. 
If we again restrict ourselves to
the interaction of the form $c\lambda \delta\Phi\phi^4/(4M_{{\rm Pl}})$, 
moduli are produced through the scatterings of the inflaton quanta
and -- at the end of the resonance stage -- $\langle \Phi^2\rangle\sim
10^{-13}\;M^2_{{\rm Pl}}$.

The analytic estimate in eq.~(\ref{es}) well reproduces this numerical result
for $\langle \Phi^2\rangle$.
The corresponding  ratio $\frac{n_{\Phi}}{s}$, derived from 
eq.~(\ref{nphis}), violates the bound (1) by seven orders of magnitude
for $c=1$.

\begin{figure}
\begin{center}
\leavevmode\epsfxsize=5.5in\epsfbox{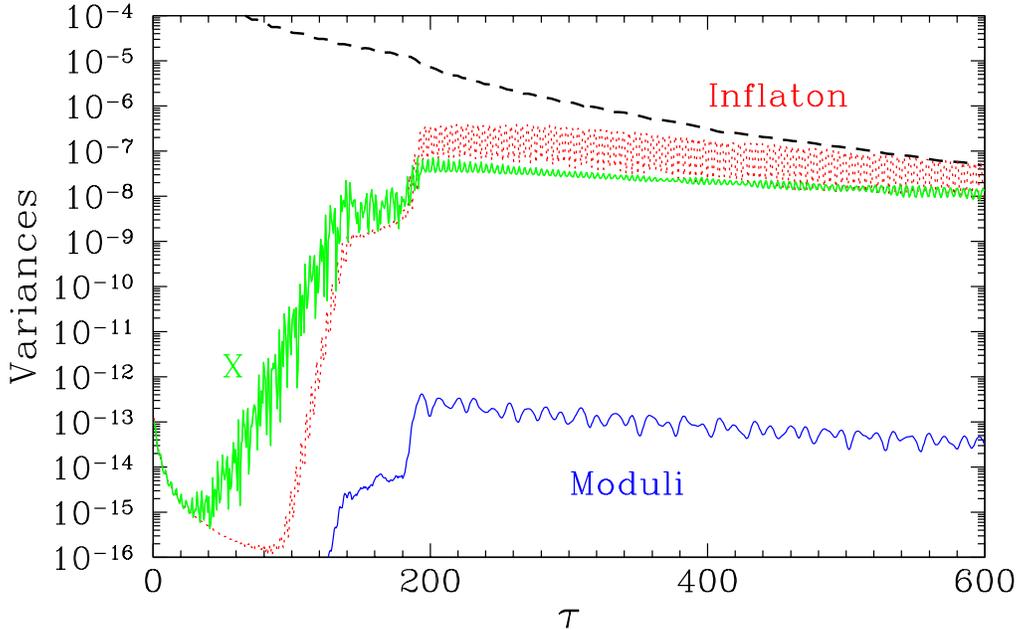}
\end{center}
\caption{The variances of the inflaton, $X$ and moduli fields in units of
$M_{\rm Pl}^2$ in a model with potential $V=\frac{\lambda}{4}
\phi^4+\frac{g^2}{2}\phi^2X^2$ and moduli interaction $g^2 
\phi^2X^2\delta \Phi /(2M_{\rm Pl})$,
with $\lambda=10^{-12}$, $q\equiv g^2/(4\lambda)=25$.
The dashed line shows the
amplitude square of the inflaton field 
oscillations in the same units.}
\label{fig:c2X}
\end{figure}

Another option is to have a coupling of the form 
$g^2\delta\Phi\phi^2X^2/(2M_{{\rm Pl}})$. In such a case
the production of moduli  
cannot take place through oscillations at the initial stage of the
parametric
resonance, 
but only via inverse decays of already created fluctuations of the $X$ field. 
This is equivalent to say 
that the effect in the Green function method is not linear in the
fluctuations, but at most quadratic. However, moduli are produced at the 
initial stage of the semiclassical thermalization 
and the result is not much different from the
previous case, see Fig.~\ref{fig:c2X}.

The main difference during the
evolution arises because the effective couplings in the system depend upon
the avereage value of the fluctuations of the moduli field. 
This
changes effectively the parameter $q$ and the  time development of the
resonance which 
is very sensitive to any change of $q$. 

We can now use 
the Green-function method to give an estimate which approximately
reproduces the results found in Fig.~\ref{fig:c2X}. The Fourier modes
of the moduli fields at time $t$ can be written as
\begin{equation}
 \Phi_{{\bf k}}(t)=\frac{cg^2}{2M_{{\rm Pl}}}\int_0^t\;
dt^\prime\;\frac{\sin\left[\omega_k(t-t^\prime)\right]}{\omega_k}
\phi_0^2(t^\prime)\;\int d^3pX_{\bf p}(t') X_{\bf k-p}(t').
\label{phic}
\end{equation}
As expected, the moduli Fourier modes are quadratic in the fluctuations
of the $X$ field. We can perform the time integral by assuming that
the leading time dependence comes from the exponential growth of
$X_{\bf p}$, which is proportional to $e^{\mu_X t}$, with $\mu_X
\simeq 0.1 \sqrt{\lambda} \phi_0(0)$. Since the momentum distribution
of the $X$ fluctuations is sharply peaked, the integral over $\bf k$
of the square of eq.~(\ref{phic}) at the end of the resonance gives
\begin{equation}
\label{es1}
\langle\Phi^2\rangle\simeq \left[\frac{g^2\phi_0^2(t_r)\langle X^2\rangle}
{4M_{\rm Pl}\mu_X\omega_k(t_r)}\right]^2\simeq 10^3 q^2
\frac{\phi_0^2(t_r)\langle X^2\rangle^2}{M^4_{\rm Pl}}.
\end{equation}

Here we have used $\omega_k\sim \sqrt{\lambda}\phi_0$
as the typical frequency.
{}From Fig.~\ref{fig:c2X} we gather that, at the maximum,
$\langle X^2\rangle\simeq 10^{-7}M^2_{{\rm Pl}}$, corrisponding
to $\phi_0^2(t_r)\simeq  10^{-5}M^2_{{\rm Pl}}$. For
$q=25$, eq.~(\ref{es1}) gives $\langle\Phi^2\rangle \simeq
10^{-13}M^2_{{\rm Pl}}$, in fair  agreement with our numerical
results.

\begin{figure}
\begin{center}
\leavevmode\epsfxsize=5.in\epsfbox{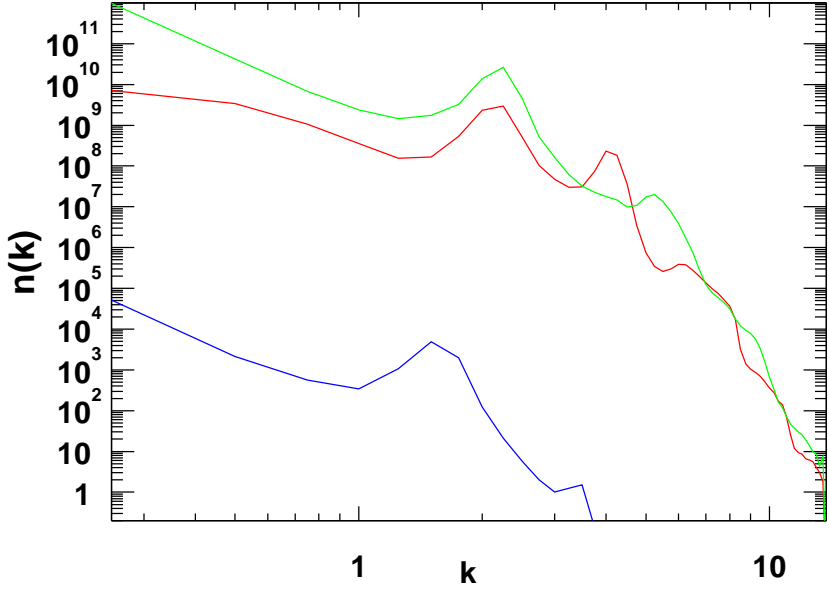}
\end{center}
\caption{Particle number densities of $X$, inflaton and moduli fields,
from top to bottom,
at $\tau = 180$ 
in the comoving reference frame, in units of $\sqrt{\lambda}\phi_0(0)$,
for the same model studied in Fig.~\ref{fig:c2X}.}
\label{fig:nofk}
\end{figure}

In all cases we also have calculated the particle occupation numbers 
to make sure that the lattice spacing and the
integration box are suitably chosen and the relevant particle momenta are well
represented. Particle number densities in the comoving reference frame are
shown in Fig.~\ref{fig:nofk} for the same model which is represented
in Fig.~\ref{fig:c2X} and at the time $\tau = 180$.  Let us also note that 
the production of moduli during the
preheating stage should be analyzed by letting the moduli be coupled  
to the matter fields in all the interaction terms. Therefore, our estimates provide a
lower limit on the abundance of moduli states.

Reassured  by the capability of reproducing the numerical results  by
analytical means, we have 
also considered   the quadratic inflaton
model with quartic coupling to another field $X$, $V=\frac{1}{2}
m^2\phi^2+ \frac{1}{2}g^2\phi^2 X^2$. The $q$-parameter is 
$q=g^2\phi_0^2(0)/4m^2$. It is also useful to introduce the
redshifted resonance parameter at the end of the resonance stage
$q_r=q(\phi_0(t_r)/\phi_0(0))^2$. Parametric resonance can fully
develop in an expanding Universe if $q_r\simgt 1$ \cite{th2}.
Resonant production is most effective for fluctuations of fields
which couple not too weakly but also not too strongly, those
with $q_r\sim 1$. For $q_r\gg 1$, the maximal size
of $X$ fluctuations is suppressed by non-linear effects
\cite{th2,th3}. The zero mode rapidly drops at the end of the
chaotization stage and all its energy is transferred to
fluctuations \cite{th3}; the variance of the inflaton field
is much larger than the value of the zero mode at the
chaotization phase.

We now want to perform an analytical estimate  of the number density
of moduli fields for moderate $q_r$. 
If the mass term has a dependence
on the modulus field $\Phi$, $m^2=m^2(\Phi/M_{{\rm Pl}})$, this will
introduce a trilinear coupling $m^2\phi^2\delta\Phi/(2M_{{\rm Pl}})$ 
between the
quanta of the modulus and the inflaton quanta.  In such a case, the 
Green-function method previously illustrated gives a modulus variance
at the end of resonance equal to 
\begin{equation}
\langle\Phi^2\rangle \simeq\frac{m^4\langle\phi^2\rangle^2}{16~
\omega^2(t_r)\mu^2M^2_{\rm Pl}} \simeq \frac{\langle\phi^2\rangle^2}
{M^2_{\rm Pl}}.
\label{quadd}
\end{equation}
Here we have taken $\omega\sim m$ as the typical energy and
$\mu\sim 0.2 \;m$~\cite{pre}. 
For $q_r\sim 1$, which corresponds to $q\sim 10^4$
\cite{th2}, the inflaton completely decays into fluctuations during the
chaotization phase; at the end of it $\phi_0^2\sim \langle \phi^2 \rangle \sim
n_\phi/m\sim 10^{-6} M^2_{{\rm Pl}}$. Therefore, from eq.~(\ref{quadd}),
we obtain $n_\Phi/n_\phi\sim
10^{-6}$.  

Notice that we could obtain the same estimate using
the Boltzmann equation for the number density $n_\Phi$:
\begin{equation} 
\dot{n}_\Phi +3H n_\Phi\sim
\frac{m^4}{\omega^4}\frac{n^2_\phi}{M^2_{{\rm Pl}}}. \end{equation}
Assuming again $\omega\sim m$ and knowing that $n_\phi$ grows
exponentially with rate $2\mu$, we get $n_\Phi/n_\phi\sim n_\phi/(\mu\;
M^2_{{\rm Pl}})$. 

At the end of the reheating stage, the energy density stored in the
inflaton is converted into a relativistic thermal bath with temperature
$T_r \sim (mn_\phi)^{1/4}$. The number
density of moduli per entropy density will then be 
\begin{equation} 
\frac{n_\Phi}{s}\sim
\frac{n_\Phi}{n_\phi}\frac{T_r}{m}\sim 10^{-6}\frac{T_r}{m}.
\end{equation} 
Since the COBE normalization gives $m\sim 10^{13}$ GeV, 
the upper bound on the reheating
temperature becomes 
\begin{equation} 
\label{m}
T_r\simlt 10^{7}\;{\rm GeV}.
\end{equation} 

If the coupling of the modulus is $g^2\delta\Phi\phi^2X^2 /(2M_{{\rm
Pl}})$, the Green function method gives
\begin{equation}
\langle\Phi^2\rangle \simeq
\left[ \frac{g^2\phi_0^2(t_r)\langle X^2\rangle}
{4M_{\rm Pl}\mu_X \omega(t_r)}\right]^2\simeq
\left[ \frac{q_r\langle X^2\rangle}{0.2M_{\rm Pl}}\right]^2.
\end{equation} 
Here we have assumed $\omega \sim m$ and $\mu_X\sim 0.2 m$.
Repeating the same argument used above, we find
$n_\Phi/s\sim 10^7q_r^2\langle X^2\rangle^2T_r/(M_{\rm Pl}^4m)$.
Since at the end of the resonance phase $q_r\sim 1$, we find that
the bound (1) gives a limit on the reheating temperature of
$10^5$~GeV  if -- for instance -- 
we use $q=10^6$ and the maximum value of $\langle X^2\rangle\sim
10^{-6}M^2_{{\rm
Pl}}$ \cite{th3}. 

So far we have always considered moduli couplings 
arising from the scalar potential, but similar results are also obtained
when the moduli couplings are derived from non-minimal kinetic terms.
For instance, from an interaction of the form $\delta \Phi {\dot \phi}^2/
M_{\rm Pl}$ we estimate
\begin{equation}
\langle\Phi^2\rangle \sim \frac{{\dot \phi}_0^2 \langle {\dot \phi}^2
\rangle}{M_{\rm Pl}^2\omega^4}\sim \frac{ \phi_0^2 \langle \phi^2
\rangle}{M_{\rm Pl}^2}.
\label{rasputin}
\end{equation}
In the case of chaotic inflation with quadratic potential, 
eq.~(\ref{rasputin}) is
equivalent to eq.~(\ref{quadd}), valid for moduli coupled
to the inflaton mass term.

\section{Moduli Production at Preheating in Hybrid Inflation}

In this section we study the moduli production
during the preheating phase following a period of hybrid inflation 
\cite{hybrid}. This scenario 
involves  two scalar fields, the inflaton field $\phi$, and
the symmetry breaking field $\sigma$, and different mass scales and 
couplings. During inflation, the inflaton field $\phi$ rolls down along a flat
potential while the Higgs field $\sigma$ is stuck at the origin providing
the
vacuum energy density driving inflation. However, when $\phi$ gets smaller
than a critical value $\phi_c$, both fields roll down very
quickly towards their present minima, completing the inflationary phase.

We take the 
simplest hybrid inflation potential as suggested by Linde \cite{hybrid}
\footnote{For other hybrid inflation models, including those motivated
by supersymmetry, see ref.~\cite{lr}.}

\begin{equation}
\label{Linde_potential}
V(\phi,\sigma) = \frac{\lambda}{4} \left(v^2 -  
\sigma^2\right)
^2
+ \frac{1}{2} m_\phi^2 \phi^2 + \frac{1}{2} g^2 \phi^2 \sigma^2\ .
\end{equation}
This potential has a valley of minima at $\sigma = 0$ for $\phi^2 >
\phi^2_c
\equiv \frac{\lambda}{g^2} v^2$.  Most of the
inflation occurs
while
$\phi$
is slowly rolling down from its initial value to $\phi_c$.  During this phase
$\sigma=0$ and the potential reduces to
$V=\frac{\lambda}{4} v^4+
\frac{1}{2} m_\phi^2 \phi^2$.
For 
$\phi < \phi_c$, the concavity of the potential in the $\sigma$ direction
changes so that the $\phi$-axis is a ridge of unstable maxima.

Inflation ends because of the growth of quantum fluctuations in $\sigma$.
For $\phi > \phi_c$, these are quickly damped out, but beneath $\phi_c$
they cause the fields to fall very quickly to global
minima at $\phi = 0$ and  $\sigma^2 = v^2$.
There, they will execute damped oscillations.

\begin{figure}
\begin{center}
\leavevmode\epsfxsize=5.5in\epsfbox{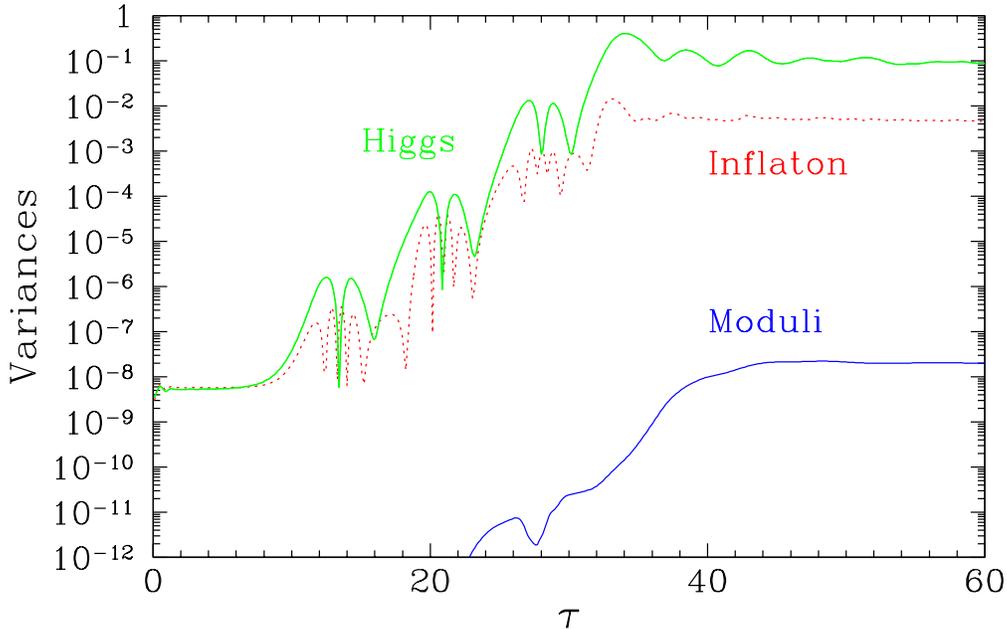}
\end{center}
\caption{Variances of the inflaton, Higgs and moduli fields in units of
$v^2$ in the hybrid model with $g^2/\lambda = 10$, $\lambda =10^{-8}$ 
and $v = 10^{-3} M_{\rm Pl}$.
The moduli are coupled according to the $g^2 \phi^2 X^2\Phi/(2M_{\rm Pl})$ 
interaction. 
}
\label{fig:hybrid}
\end{figure}

The results of numerical integration for the case $v= 10^{-3} M_{\rm
Pl}$
are shown in Fig. \ref{fig:hybrid}, where we have
supposed that the modulus is coupled to the inflaton field and the Higgs
field via the coupling $g^2\Phi \phi^2
\sigma^2/(2M_{{\rm Pl}})$. 
In hybrid models the
coupling constant
$\lambda$ does
not have to obey the same restrictions as in the case of chaotic
inflation. However, for the numerical integration we have chosen a rather
low value, $\lambda = 10^{-8}$. This is because the particle number
in the inflaton or Higgs excitations 
at the beginning of chaotization is $n(k) \sim 1/\lambda$ and  to be
able to
treat moduli fields classically, we need occupation numbers 
$n_{\Phi} \gg 1$. Since $n_{\Phi} \ll n(k)$, we need a rather
small value of $\lambda$. For the run presented in Fig. \ref{fig:hybrid}
we have chosen $g^2/\lambda = 10$. 

The initial conditions were chosen
in such a way that Higgs oscillations occur only around one of the minima,
and domain creation and other effects leading to complete chaotisation during
the first oscillation (see Ref. \cite{Hybrid}) can be disregarded.

Our numerical findings  show 
that the variance of the moduli fields does depend only upon the ratio
$g^2/\lambda$.

The Green function method tells us that 
\begin{equation}
\label{eshybrid}
\langle\Phi^2\rangle\simeq \frac{g^4}{M^2_{{\rm Pl}}}
\frac{\langle\phi^2\rangle^2}{\omega_\phi^2}
\frac{\langle\sigma^2\rangle^2}{\omega_\sigma^2}
\simeq \frac{g^2}{\lambda}\frac{\langle\phi^2\rangle^2}{M^2_{\rm Pl}}
,
\end{equation}
since
the typical
frequencies are given by  
$\omega^2_\phi\sim
\lambda \langle \sigma \rangle^2$ and $\omega_\sigma^2\sim
g^2 \langle \sigma \rangle^2$. 
Using the same values of the couplings and mass scales adopted
in Fig. 5 and reading from there the variance of the inflaton field
$\langle \phi^2\rangle \sim 10^{-2}v^2$, we get
$\langle\Phi^2\rangle\sim 10^{-9} v^2$, in fair
agreement
with our numerical results. Notice that the numerical values of the
inflaton and Higgs variances at their maximum are well reproduced by 
$\langle\phi^2\rangle\sim (\lambda/g^2)v^2$
and $\langle\sigma^2\rangle \sim  v^2$. Using these values we estimate
$\langle\Phi^2\rangle\sim (\lambda/g^2)(v^4/M^2_{{\rm Pl}})$,
in good agreement with our numerical results and with the  fact that
these results depend only upon the
ratio $g^2/\lambda$.

If the final reheating takes place via decay of the
lightest particles in the system, {\it i.e.} the Higgs particles
with mass squared $\lambda v^2$, the final moduli number
density-to-entropy-ratio is given by
\begin{equation}
\frac{n_\Phi}{s}\sim \frac{n_\Phi}{n_\sigma}\frac{T_r}{\sqrt{\lambda}v}
\end{equation}
which implies
\begin{equation}
\label{hybb}
T_r\simlt \sqrt{\lambda} \;
\left( \frac{10^{-3}~M_{\rm Pl}}{v}\right)
\left( \frac{g^2/\lambda}{10}\right)\; 10^{11}~{\rm GeV}.
\end{equation}
 This result illustrates how, in certain ranges of parameters,
the coherent
production of moduli can become more dangerous than the generation of
moduli via thermal scatterings.

\section{Modulino Production at Preheating}

So far we have been considering the
generation of scalar moduli during the preheating stage. 
In this section we consider the production of the fermionic partners 
$\psi_\Phi$ of the
moduli fields, sometimes dubbed modulini. As for the scalar moduli, the 
modulini are generically coupled to the matter fields through couplings
suppressed by $M_{{\rm Pl}}$. 

Consider -- for instance --  the superpotential
$W=\frac{\sqrt{\lambda}}{3}\phi^3$ giving rise to the inflaton potential
$V=\lambda\phi^4$. If the coupling $\lambda$ is a function of $\Phi/
M_{{\rm Pl}}$, where now $\Phi$ has to be considered as a superfield, the
Lagrangian will contain the following coupling
\begin{equation}
\frac{\sqrt{\lambda}}{M_{{\rm Pl}}}\overline{\psi}_\Phi\psi_\phi\phi^2,
\end{equation}
where  have denoted with $\psi_\phi$ the inflatino, the 
fermionic partner of the inflaton field. 

We can again make use  of the  
Green function method to obtain the Fourier component of $\psi_\Phi$. 
Using the retarded Green function for a fermionic field, we find
\begin{equation}
\label{greenfermion}
\psi_\Phi({{\bf k}},t)=\frac{\sqrt{\lambda}}{M_{{\rm Pl}}}\int_0^t\;
dt^\prime\;\left\{
-i\gamma^0\sin\left[\omega_k(t-t^\prime)\right]
-\frac{{\vec{\gamma}}\cdot{\vec{k}}}{\omega_k}
\cos\left[\omega_k(t-t^\prime)\right] \right\}
\;
\phi_0^2(t^\prime)\;
\psi_\phi({{\bf k}},t^\prime).
\end{equation}
We can then approximatly estimate the 
quantity $\langle \overline{\psi}_\Phi \psi_\Phi
\rangle$ as
\begin{equation}
\label{pit}
\langle \overline{\psi}_\Phi\psi_\Phi\rangle\simeq 
\frac{\lambda}{M^2_{{\rm Pl}}}\frac{\phi_0^4}{\omega^2}
\langle \overline{\psi}_\phi\psi_\phi\rangle.
\end{equation}
One
can now use as a guide  the recent results obtained in the theory of generation of
Dirac fermions during and after inflation \cite{ferm}.
During the first inflaton
oscillations, the Fermi distribution function of the inflatini 
is rapidly saturated up to some
maximum value of the momentum $k$, {\it i.e} $n_k\simeq  1$ for $k\simlt
k_{{\rm max}}$ and it is zero otherwise.  The resulting number density is
therefore $n_{k}\sim k_{{\rm max}}^3$. The value of $k_{{\rm max}}$ is
expected to be roughly of the order of the inverse of the time-scale needed for
the change of the mass  of the inflatini. In the model
under consideration, such a mass  
changes by an amount $\sqrt{\lambda}\phi_0$ in a 
time scale $(\sqrt{\lambda}\phi_0)^{-1}$ and  one expects 
$k_{{\rm max}}\sim \sqrt{\lambda}\phi_0$. 
Therefore
\begin{equation}
\langle \overline{\psi}_\phi\psi_\phi\rangle\sim
\lambda^{3/2}\phi^3_0(0),
\end{equation}
where we have used the fact that most of 
the fermion production takes
place during the  first oscillation. Under these circumstances, one can
also assume that the number density of inflatini $n_{\psi_\phi}$ 
and modulini $n_{\psi_\Phi}$ is well measured by 
$\langle \overline{\psi}_\phi\psi_\phi\rangle$
and $\langle \overline{\psi}_\Phi\psi_\Phi\rangle$, respectively. This gives
\begin{equation}
\frac{n_{\psi_\Phi}}{n_{\psi_\phi}}\sim 
\frac{\phi^2_0(0)}{M^2_{{\rm Pl}}},
\end{equation}
where we have taken $\omega\sim k_{{\rm max}}$ in Eq. (\ref{pit}).

Finally, the ratio of the number density of modulini 
in units of the entropy density $s\sim\rho^{3/4}\sim \lambda^{3/4}\phi_0^3$ 
is given by
\begin{equation}
\frac{n_{\psi_\Phi}}{s}\sim 
\left(\frac{\lambda}{10^{-13}}\right)^{3/4}
\left(\frac{\phi_0(0)}{0.3~M^2_{\rm Pl}}\right)^2 10^{-11}.
\end{equation}
This result will contradict 
the bound (\ref{lll}) by about an order of magnitude
when the energy density in the scalar field 
is transferred to the energy density of a hot gas of relativistic particles.

Similar considerations apply to the inflaton model with quadratic
potential. Suppose there
is a coupling of the form $m\phi^2\Phi/M_{{\rm Pl}}$ in the superpotential.
This gives rise in the Lagrangian 
to a coupling of the modulini with the fermionic partners
of the inflaton of the form $\frac{m}{M_{{\rm Pl}}}
\overline{\psi}_\Phi\psi_\phi\phi$. 
Using the Green function method we again obtain $n_{\psi_\Phi}\sim 
n_{\psi_\phi} (\phi_0(0)/M_{{\rm Pl}})^2$. This gives $n_{\psi_\Phi}/
n_{\psi_\phi}\sim
(m/M_{{\rm Pl}})^2\sim 10^{-12}$ and the generation of dangerous relics is
not large enough to
provide a strong constraint on the reheating temperature.

In the case of the hybrid model though, the bound is more
restrictive. In supersymmetric hybrid models, the coupling constants
$\lambda$ and $g$ are related, and typically one finds $\lambda /g^2 \sim 
1$~\cite{lr}.
The COBE normalization  implies that typically $v\simeq 5\times 10^{15}$ GeV.
During the oscillations of the
Higgs and the inflaton field, one expects that the number density of
``Higgsini '' $\psi_\sigma$ and inflatini $\psi_\phi$ is of the order of the number density of Higgs 
particles excited during the preheating phase, 
$n_{\psi_\sigma}\sim n_{\psi_\phi}
\sim n_\sigma\sim \lambda^{3/2} v^3$. If the modulus couples to the Higgs
superfield via the interaction $\sqrt{\lambda}(\Phi/M_{{\rm Pl}})
\sigma^2\phi$ in the superpotential, the number density of modulini
is expected to be 
\begin{equation}
\frac{n_{\psi_\Phi}}{n_\sigma}\sim \left(\frac{v}{M_{{\rm Pl}}}\right)^2.
\end{equation}
The corresponding bound on the reheating temperature can be obtained using the
same arguments which led us to eq. (\ref{hybb}); we find 
$T_r \simlt \sqrt{\lambda} \;
\left(5\times 10^{15} {\rm GeV} /v\right)\; 10^{10}{\rm GeV}$.

\section{Conclusions}

In this paper we have investigated the production of scalar and fermionic moduli fields 
which occurs during the preheating stage after inflation. If
produced with too large abundances, the late decays of these relics may jeopardize the successful predictions
of standard big-bang nucleosynthesis. Since the moduli couple to any form of matter at least
gravitationally, their generation is unescapable in the large class of inflationary models which
have a preheating stage. They are generated as coherent states initially.
{}From this point of view, the
production mechanism of moduli  differs from the generation mechanism of squeezed states during
the parametric resonance which has been studied in  recent years. 
We have shown that moduli are efficiently produced during the preheating stage. The corresponding upper
bound on the reheating temperature is model dependent and sensitive to the 
details of the stage of thermalization
following the period of parametric resonance. It is 
often tighter than the bound of $10^9$ GeV found considering the generation of
moduli via thermal scatterings during reheating.


\end{document}